\documentclass[conference]{IEEEtran}
\IEEEoverridecommandlockouts

\usepackage{cite}
\usepackage{amsmath,amssymb,amsfonts}
\usepackage{algorithmic}
\usepackage{graphicx}
\usepackage{textcomp}
\usepackage{xcolor}
\usepackage{enumitem}
\usepackage{mathtools}
\PassOptionsToPackage{hyphens}{url}\usepackage[hidelinks]{hyperref}
\usepackage{tikz,pgfplots}
\usepackage{grffile}
\pgfplotsset{compat=newest}
\usepgfplotslibrary{patchplots}
\usetikzlibrary{backgrounds,plotmarks,fit,positioning,arrows,shapes,shapes.multipart,calc,arrows.meta}
\usetikzlibrary{fit,positioning,arrows,shapes,shapes.multipart,calc,arrows.meta,shapes.geometric,shapes.misc}
\usetikzlibrary{decorations.pathreplacing,angles,quotes,calligraphy}
\usetikzlibrary{automata}
\newlength{\rx}
\newlength{\ry}

\allowdisplaybreaks

\usepackage{subfigure}

\usepackage{vmr-symbols-vecbold}
\usepackage{standard-macros}

\usepackage[toc,acronym,nonumberlist,nopostdot]{glossaries}
\newacronym{IoT}{IoT}{Internet of Things}
\newacronym{MAC}{MAC}{multiple access channel}
\newacronym{UMA}{UMA}{unsourced multiple access}
\newacronym{TUMA}{TUMA}{type-based unsourced multiple access}
\newacronym{TBMA}{TBMA}{type-based multiple access}
\newacronym{AWGN}{AWGN}{additive white Gaussian noise}
\newacronym{EP}{EP}{expectation propagation}
\newacronym{AMP}{AMP}{approximate message passing}
\newacronym{PMF}{PMF}{probability mass function}
\newacronym{PDF}{PDF}{probability density function}
\newacronym{SNR}{SNR}{signal-to-noise ratio}

\renewcommand{\define}{=}

\newcommand{\n}{{\uN}} 
\newcommand{\Ka}{{\uK_{\rm a}}} 
\newcommand{\Ma}{{\M_{\rm a}}} 
\newcommand{\estMa}{{\west{\M}_{\rm a}}} 
\newcommand{\power}{\uP} 
\newcommand{\M}{\uM} 

\newcommand{\quant}{{\sf quant}}
\newcommand{\enc}{{\sf enc}}
\newcommand{\dec}{{\sf dec}}

\newcommand{\WS}[2]{\ensuremath{{\opW}_{p}\lefto(#1,#2\right)}} 
\newcommand{\dTV}{\ensuremath{\overline{\opT\opV}}}
\newcommand{\dWS}{\ensuremath{\overline{\opW}}}

\newcommand{\Binpmf}[3]{{\rm Bin}\lefto(#1;#2,#3\right)}

\newcommand{\of}[1]{^{(#1)}}

\def\BibTeX{{\rm B\kern-.05em{\sc i\kern-.025em b}\kern-.08em
    T\kern-.1667em\lower.7ex\hbox{E}\kern-.125emX}}
    
\begin{document}

\title{Type-Based Unsourced Multiple Access  \vspace{-.3cm}
	\thanks{This work was supported in part by the Swedish Research Council under grants 2021-04970 and 2022-04471, and by the Swedish Foundation for Strategic Research.}
}

\author{
\IEEEauthorblockN{Khac-Hoang Ngo, Deekshith Pathayappilly Krishnan, Kaan Okumus, Giuseppe Durisi, and Erik G. Str\"om}

 \IEEEauthorblockA{Department of Electrical Engineering, Chalmers University of Technology, 41296 Gothenburg, Sweden \vspace{-.5cm}}
}

\maketitle

\begin{abstract}
    We generalize the type-based multiple access framework proposed by Mergen and Tong (2006) to the case of unsourced multiple access. In the proposed framework, 
    each device tracks the state of a physical/digital process, quantizes this state, and communicates it to a common receiver through a shared channel in an uncoordinated manner. The receiver aims to estimate the type of the states, i.e., the set of states and their multiplicity in the sequence of states reported by all devices. We measure the type estimation error using the Wasserstein distance. Considering an example of multi-target position tracking, we show that type estimation can be performed effectively via approximate message passing. Furthermore, we determine the quantization resolution that minimizes the type estimation error by balancing quantization distortion and communication error. 
\end{abstract}



\section{Introduction} \label{sec:intro}
Future 
\gls{IoT} services 
are enabled by the capability to collect data from a massive number of low-cost distributed 
devices.\footnote{In this paper, we shall refer to devices also as users or sensors.} 
These devices 
transmit short packets in a sporadic and uncoordinated manner. 
To address this scenario, Polyanskiy proposed the \gls{UMA} framework~\cite{PolyanskiyISIT2017massive_random_access}, where all users employ a common codebook and the receiver returns an unordered list of messages. 
This 
work ignited an active research area that aims to characterize theoretical limits on the achievable energy efficiency~\cite{Kowshik2020,Ngo2023_randomActivity} and devise coding schemes approaching these limits~\cite{fengler2019sparcs,Amalladinne2022}.


In \gls{UMA}, the users generate their messages independently and 
the receiver 
treats any message collision between users 
as an error. 
However, real-world scenarios often involve messages describing physical/digital processes correlated across users. Moreover, the receiver may be interested not only in the set of transmitted messages but also in their frequency of occurrence. For instance, in multi-target tracking~\cite{Hoffman2004,Schuhmacher2008,Rahmathullah2017}, each message may represent the state of a target, and the receiver may seek to estimate the number of sensors tracking each target. In federated learning, 
each message may correspond to a quantized local model update. To aggregate these updates, the server needs to estimate the number of users submitting each quantized update~\cite{Qiao2024massive,Gunduz2003BITS}. In these contexts, the users sense/sample parameters from physical/digital processes, {\em quantize} these parameters, and {\em communicate} them to the receiver. The \gls{UMA} framework focuses only on the communication phase.

The problem of estimating the type, i.e., the number of users sending each message, 
was formulated in~\cite{Mergen2006} under the name of~\gls{TBMA}. 
However, the authors assume that each message is assigned an orthogonal codeword, which entails a large latency if the number of messages is large. 
Recent research has explored type estimation in random-access scenarios with nonorthogonal codewords. In particular, \cite{Dommel2021} considers \gls{IoT} monitoring over a fog radio access network,~\cite{Zhu2023} proposes an information-bottleneck-based codebook design and a neural-network-based decoder, and~\cite{Qiao2024massive} addresses the federated learning setup. However, these works consider small systems~\cite{Dommel2021,Zhu2023}. Hence, it is unclear whether the proposed solutions can be applied to the massive access regime. Moreover, they consider a specific application~\cite{Zhu2023,Qiao2024massive} rather than developing a general framework. 

In this paper, we present \gls{TUMA}, a setup that bridges \gls{TBMA} and \gls{UMA}, propose a general performance metric 
for this setup, and benchmark under this metric three algorithms to estimate the type. 
In \gls{TUMA}, we assume that a large number of users track the states of multiple processes and report these states to a receiver. Each user tracks a process and represents the state of this process by a message that is used to i) map the state to a quantized state belonging to a quantization codebook, and ii) select an element of a communication codebook. Both the quantization and communication codebooks are the same for all users. To report the state, the user sends the corresponding communication codeword in an uncoordinated manner. 
The receiver aims to estimate 
the type of the states, i.e., the set of distinct states and their multiplicity. 
To assess the type estimation error, we use the Wasserstein distance, a metric commonly used in optimal transport~\cite{Payre2019_optTransport}. 

We compare three existing type estimators: the \gls{AMP} estimator derived for \gls{UMA} in~\cite{fengler2019sparcs}, the scalar \gls{AMP} estimator derived in~\cite[Sec. IV-C]{Meng2021} and adopted in~\cite{Qiao2024massive}, and the \gls{EP} estimator derived in~\cite[Sec.~IV-D]{Meng2021}. The \gls{AMP} estimator turns out to have the same complexity order as scalar \gls{AMP} and a lower complexity order than \gls{EP}. Via an example pertaining to multi-target position tracking, we show that \gls{AMP} also achieves a better performance than scalar \gls{AMP} and \gls{EP}. Our numerical results highlight a trade-off between quantization and communication: increasing the quantization resolution 
reduces the distortion but degrades the communication performance. 
The average Wasserstein distance 
is minimized at a certain quantization resolution. Remarkably, the optimal resolution cannot be achieved using orthogonal codebooks as in~\gls{TBMA}~\cite{Mergen2006}.

\subsubsection*{Notation}
We denote system parameters by uppercase non-italic
letters, e.g.,~$\uK$. Uppercase italic letters, e.g.,~$X$, denote scalar random variables and their realizations are in lowercase, e.g.,~$x$. Vectors are denoted likewise in boldface, e.g., a random vector~$\rvecx$ and its realization~$\vecx$. 
We denote the $n \times n$ identity matrix by~$\matidentity_n$. 
The superscript~$\tp{}$ 
stands for transposition. 
We denote sets with calligraphic letters, e.g., $\setS$, 
the Gaussian vector distribution with mean $\vecmu$ and covariance matrix $\matA$ by $\normal(\vecmu, \matA)$, and its \gls{PDF} by $\normal(\cdot; \vecmu, \matA)$. We denote the \gls{PMF} of a binomial distribution with parameters $(n,p)$ 
 by $\Binpmf{\cdot}{n}{p}$. 
We use $\vecnorm{\cdot}$ to denote the $\ell^2$-norm; 
$[n] \define \set{1,\dots,n}$; $\propto$ means ``proportional to''. 
We define a discrete measure with weights $p_1,\dots,p_n$ (with $\sumo{i}{n}p_i = 1$) and locations $\vecx_1,\dots,\vecx_n \in \setX$ as $\sumo{i}{n}p_i \dirac(\vecx_i)$, where $\dirac(\cdot)$ is the Dirac measure. The set of all discrete measures defined on~$\setX$ is denoted by $\setP(\setX)$. 

\subsubsection*{Reproducible Research}
The Matlab code 
is available at: \url{https://github.com/khachoang1412/TUMA}. 

\section{System Model} \label{sec:model}
We consider a setting in which a large number of sensors track the states of $\Ma$ targets and report these states, in an {uncoordinated manner}, to a receiver over $\n$ uses of a stationary Gaussian multiple access channel. At a given time, $\Ka$ sensors are active, i.e., they see a target. In a typical \gls{IoT} setting, $\Ka$, $\Ma$, and $\n$ are in the order of $10^2$--$10^3$, $10^1$--$10^2$, and $10^3$--$10^4$, respectively\textemdash see~\cite{PolyanskiyISIT2017massive_random_access} and \cite[Rem.~3]{Zadik2019}. For concreteness, we will focus on a multi-target tracking scenario and use terms such as ``sensors'', ``targets'', and ``states''. However, as we shall clarify at the end of this section, the setup is general. 
Let $\rvecx_k \in \reals^\n$ be the signal transmitted by sensor~$k$. The corresponding 
received signal is given by 
\begin{equation} \label{eq:received_signal}
    \rvecy = \textstyle\sum_{k=1}^{\Ka} \rvecx_k + \rvecz
\end{equation}
where $\rvecz \sim \normal(\veczero,\matidentity_\n)$ is the Gaussian noise, which is independent of the transmitted signals. The transmitted signals satisfy the power constraint $\vecnorm{\rvecx_k}^2 \le \n\power$, $\forall k \in [\Ka]$. Here, $\power$ is also identified with the \gls{SNR}.

We now describe how the transmitted signals are generated. Each target $i \in [\Ma]$ has a state $\rvecs_i$ (e.g., its position) generated from a state space~$\setS$ according to a distribution~$p_\rvecs$. 
We assume that each sensor~$k$ reports the state of a single target~$O_k$, and that this target is selected according to a distribution~$p_{O_k}$. Specifically, we assume that the sensor maps $\rvecs_{O_k}$ to one of the quantized states in a quantization codebook $\setQ = \set{\vecq_1, \dots, \vecq_\M}$. Let $W_k \in [\M]$ be the index of the quantized state associated to $\rvecs_{O_k}$. To report $W_k$ to the receiver, sensor~$k$ transmits $\rvecx_k = \sqrt{\n\power} \vecc_{W_k}$ where $\vecc_{W_k}$ belongs to a communication codebook $\setC = \set{\vecc_1, \dots, \vecc_\M}$ with $\vecnorm{\vecc_i} \le 1$, $\forall i \in [\M]$. All sensors employ the same quantization codebook $\setQ$ and communication codebook $\setC$. The receiver aims to estimate the discrete measure $\sumo{i}{\Ma} T_i \delta(\rvecs_i)$, where $T_i \in [0,1]$ is the fraction of sensors that report target~$i \in [\Ma]$, i.e., the number of sensors reporting target~$i$ is $T_i \Ka$. This measure represents the type 
of the sequence $\rvecs_{O_1}, \dots, \rvecs_{O_\Ka}$. 
%
    
    For a given quantization codebook $\setQ$ and communication codebook $\setC$, the design of a \gls{TUMA} system consists of specifying the following functions:\footnote{We assume perfect sensing, i.e., the sensors know the exact state of their tracked target. Sensing errors can be straightforwardly incorporated in our model by introducing a sensing function before the quantization function.}
    \begin{itemize}[leftmargin=*]
        \item A quantization function $\quant_\setQ \sothat \setS \to [\M]$ that produces the message $W_k = \quant_\setQ(\rvecs_{O_k})$ of sensor~$k$ associated to the state $\rvecs_{O_k}$ of the target~$O_k$ tracked by sensor~$k$. The quantized state of target $O_k$ is then $\vecq_{W_k}$.

        \item An encoding function $\enc\sothat [\M] \to \setC$ that produces the codeword $\vecc_{W_k} = \enc(W_k)$ 
        corresponding to the message~$W_k$. 

        \item A decoding function $\dec \sothat \reals^\n \to \setP(\setQ)$ that provides an estimate 
        of $\sumo{i}{\Ma} T_i \delta(\rvecs_i)$.  Specifically, the decoder first finds an estimate $\west{\rveck} = \tp{[\west{K}_1, \dots, \west{K}_\M]}$ of $\rveck = \tp{[K_1, \dots, K_\M]}$, where $K_i$ is the number of sensors that send message $i\in [\M]$. Then, it estimates the type as $\sumo{i}{\M} \frac{\west{K}_i}{\sumo{j}{\M} \west{K}_j} \delta(\vecq_{i})$. The estimated type can be written compactly as $\sumo{i}{\estMa}\west{T_i} \delta(\vecq_{\west{W}_i})$ where $\estMa = |\{i \in [\M] \sothat \west{K}_i > 0\}|$, $\big\{\west{W}_1, \dots, \west{W}_{\estMa}\big\} = \big\{i \in [\M] \sothat \west{K}_i > 0\big\}$, and $\west{T}_i = {\west{K}_{\west{W}_i}}/ \sum_{j=1}^\M \west{K}_{j}$, $i \in [\estMa]$. 
    \end{itemize}
    Hereafter, we assume that the number of targets $\Ma$ and the number of sensors $\Ka$ are known to the decoder. Note that since some targets may not be tracked by any sensor and multiple targets may be mapped to a common quantized state, we may have that $|\{i\in [\M] \sothat K_i > 0\}| < \Ma$. Therefore, the decoder may return a type with support size $\estMa < \Ma$. 
        
    To assess the type estimation error, we consider the metric
    \begin{equation} \label{eq:distance_constraint}
        \dWS_p \define \textstyle\Exop\Big[\WS{\sumo{i}{\Ma}T_i \delta(\rvecs_i)}{\sumo{i}{\estMa} \west{T_i} \delta(\vecq_{\west{W}_i})}\Big]
    \end{equation}
    where the expectation is with respect to $p_\rvecs$, $\set{p_{O_k}}_{k=1}^\Ka$, and the additive noise. Here, $\WS{\cdot}{\cdot}$ is the $p$-Wasserstein distance~\cite[Chap.~2]{Payre2019_optTransport} defined as
    \begin{align}
        &\WS{\textstyle\sumo{i}{\Ma}T_i \delta(\rvecs_i)}{\textstyle\sumo{i}{\estMa} \west{T_i} \delta(\vecq_{\west{W}_i})} \notag \\ 
        &\define \inf_{e_{i,j} \in [0,1], i\in[\Ma], j \in [\estMa]} \Bigg( \sumo{i}{\Ma} \sumo{j}{\estMa} e_{i,j} \vecnorm{\rvecs_i - \vecq_{\west{W}_j}}^p\Bigg)^{\!1/p}\! \label{eq:WS} \\
        &\qquad \text{subject to~} \textstyle\sumo{i}{\Ma} e_{i,j} = \west{T}_j, \sumo{j}{\estMa} e_{i,j} = T_i. \notag
    \end{align}
     The Wasserstein distance 
     provides a measure of similarity between distributions that captures both their shape and support. This distance and its generalizations have been widely used in multi-target tracking~\cite{Hoffman2004,Schuhmacher2008,Rahmathullah2017}.

\begin{figure}[t!]
    \centering
    \scalebox{.75}{
    \begin{tikzpicture}
        \node[rectangle, fill = blue!10, draw = blue!20, minimum height= 5.5cm, minimum width=6cm, label={[xshift=2.4cm,yshift=-.75cm]above:{{UMA}}}, node distance = 3cm] at (15.4,3.2) {};

         \node[rectangle, draw, black, minimum height = 3.5cm,minimum width = 1cm, node distance=3.7cm, align=center] (decoder) at (7.5,3.75) {$\rvecs_1$\\$\rvecs_2$ \\ $ \vdots$ \\ $\rvecs_{{\Ma}}$};

        \node[rectangle, draw, black,minimum width = 2.2cm, node distance = 3cm, align=center] (1) at (10.9,5){Quantizer $\quant_\setQ$};
        
        \node[rectangle, draw, black, minimum width = 2.2cm, node distance = 2.5cm, align=center] (2) [below of=1] {Quantizer $\quant_\setQ$};

        \draw[-latex] (8,5) -- node [pos=.6,above] {$\rvecs_{O_1}$} (1);
        \draw[-latex] (8,2.5) -- node [pos=.65,above] {$\rvecs_{O_\Ka}$} (2);



        \node[rectangle, draw, black, minimum width = 2.2cm, node distance=3.5cm, align=center] (5) [right of=1] {Encoder $\enc$};
        \node[rectangle, draw, black, minimum width = 2.2cm, node distance=3.5cm, align=center] (6) [right of=2] {Encoder $\enc$};

        \draw[-latex] (1) -- node[midway, above] {${W_1}$} (5);
        \draw[-latex] (2) -- node[midway, above] {${W_\Ka}$} (6);
        \draw (5) -- node[midway, above right] {$\vecc_{W_1}$} (16,5);
        \draw (6) -- node[midway, below right] {$\vecc_{W_\Ka}$} (16,2.5);

        \node[draw,
            circle,
            minimum size=.75cm,
            fill=blue!10, node distance = 3cm
        ] (sum) at (17,3.75){};
         
        \draw (sum.north) -- (sum.south)
            (sum.west) -- (sum.east);

        \draw[-latex] (16,5) -- (sum);
        \draw[-latex] (16,2.5) -- (sum);

        \node[rectangle, black, node distance = 3cm] at (11,4) {$\vdots$};

        \node[rectangle, black, node distance = 3cm] (noise) at (18,3.75) {$\rvecz$};
        
        \draw[-latex] (noise) -- (sum);

        \node[rectangle, draw, black, minimum width = 2.2cm, node distance=3.5cm, align=center] (decoder) at (14.5,1) {Decoder $\dec$};

        \draw (sum) -- (17,1);

        \draw[-latex] (17,1) -- (decoder);
        \node[rectangle] at (16.25,1.33){$\rvecy$};

        \node[rectangle, black, node distance = 3cm] (Kest) at (12.7,1) {$\west{\rveck}$
        };
        
        \node[rectangle, black, node distance = 3cm] (output) at (10,1) {$\sumo{i}{\estMa} \west{T}_i \delta(\vecq_{\west{W}_i})$
        };

        \draw[-latex] (decoder) -- (Kest);
        \draw[-latex] (Kest) -- (output);

        \node[rectangle, red, draw, dashed, minimum height= 1.1cm, minimum width=7.2cm, label={[xshift=-1cm]above:{Sensor $\Ka$}}, node distance = 3cm] at (12.15,2.5) {};

        \node[rectangle, red, draw, dashed, minimum height= 1.1cm, minimum width=7.2cm, label={[xshift=-1cm]above:{Sensor $1$}}, node distance = 3cm] at (12.15,5) {};
    \end{tikzpicture} 
    }
    \vspace{-.2cm}
    \caption{Illustration of the proposed \gls{TUMA} framework and its relation to \gls{UMA}, which is represented by the shaded box.}
    \label{fig:framework}
    \vspace{-.4cm}
\end{figure}

We illustrate the proposed \gls{TUMA} framework in Fig.~\ref{fig:framework}. This framework generalizes both \gls{UMA}~\cite{PolyanskiyISIT2017massive_random_access} and \gls{TBMA}~\cite{Mergen2006}. We highlight next some key differences. First, \gls{UMA} accounts for the transmission of the messages $\{W_k\}_{k=1}^\Ka$ and the support recovery of this set. That is, an \gls{UMA} system consists only of the encoder and decoder in Fig.~\ref{fig:framework}. In \gls{TUMA}, we let these messages represent the quantized states of the targets and aim to recover the type of the reported states. Second, in \gls{UMA}, the messages $\{W_k\}_{k=1}^\Ka$ are generated uniformly at random from~$[\M]$, and because $\Ka \ll \M$, the messages are likely distinct. \gls{UMA} treats message repetitions as errors. In \gls{TUMA}, we regard such repetitions as a component of the type to be estimated by the receiver. \gls{UMA} addresses the per-sensor communication error: an error occurs for sensor~$k$ if $W_k \notin \{\west{W}_i\}_{i=1}^\estMa$. In \gls{TUMA}, we address the overall type-estimation performance. 
Finally, while the original \gls{TBMA} framework~\cite{Mergen2006} 
assumes that $\n \ge 
\M$ and thus $\setC$ contains orthogonal codewords, \gls{TUMA} allows $\n$ to be smaller than $\M$, a scenario for which orthogonal codewords are not possible. 
    
    

The \gls{TUMA} framework is relevant not only for multi-target tracking but also for various other scenarios, including the following.


\subsubsection{\gls{IoT} Monitoring System}
$\Ka$ \gls{IoT} sensors monitor the status of a system and send updates to a gateway.
Let the $\Ma$ targets be $\Ma$ events that occur at a given time 
and trigger the sensors. Each event is identified by its index out of $\M$ possible events, and  
can be reported by multiple sensors. The common-alarm scenario considered in~\cite{Ngo2024_commonAlarm} is a special case of this setup. It corresponds to the case in which only one event occurs and is reported by all sensors. 

\subsubsection{Point-Cloud Transmission~\cite{Bian2023_pointCloud}}
A point cloud is a collection of three-dimensional ($3$D) data points and their associated attributes (such as color and temperature) generated by, e.g., light detection and ranging (LiDAR) sensors and depth cameras.
To represent the $3$D points, we let the state space $\setS$ be a $3$D volume. The quantization codebook defines a regular $3$D grid. Each state corresponds to a point in the grid. Different LiDAR sensors/cameras detect different points based on their angle of view, and report these points to the receiver.

\subsubsection{Digital Over-the-Air Computation~\cite{Qiao2024massive}}
In this scenario, the sensors coincide with the targets and correspond to clients in a distributed system. The states correspond to local vectors at the clients. The clients submit their vectors to a central server which computes the mean of these vectors. This is a subroutine in many distributed learning and optimization schemes, such as federated learning. Following \gls{TUMA}, each client quantizes its vector and sends the quantization index to the server. The server estimates the type of the vectors submitted by all clients and computes the mean of the  estimated type. 


\section{Decoder Design} \label{sec:decoder}
In this section, we discuss the design of decoding function~$\dec$. We rewrite the received signal~\eqref{eq:received_signal} as
\begin{equation}
    \rvecy = \textstyle \sqrt{\n \power}\sumo{i}{\M} K_i \vecc_i + \rvecz = \sqrt{\n \power} \matC \rveck + \rvecz, \label{eq:received_signal_2}
\end{equation}
where $\matC = [\vecc_1,\dots,\vecc_\M]$ is the codebook matrix.
Given a realization $\vecy$ of $\rvecy$, we aim to estimate $\rveck$. 
This is a compressed sensing problem for which many algorithms are available. For the sake of decoder design, it is convenient to decouple the $\rveck$-estimation performance from the overall type estimation performance~\eqref{eq:WS}. Specifically, we use the average total variation distance between the empirical distributions induced by $\rveck$ and $\west{\rveck}$ as the communication performance metric:
\begin{equation}
	\dTV \define \Exop\Bigg[\frac{1}{2} \sum_{i=1}^\M \Bigg|\frac{K_i}{\sum_{j=1}^\M {K}_{j}} - \frac{\west{K}_{i}}{\sum_{j=1}^\M \west{K}_{j}}\Bigg|\Bigg]. \label{eq:TV}
\end{equation}
Note that if 
$K_i \in \set{0,1}, \forall i \in [\M]$, $\dTV$ coincides with the per-sensor error probability considered in \gls{UMA}~\cite[Def.~1]{PolyanskiyISIT2017massive_random_access}.

Let $p_{K_i}(k)$ be the marginal prior distribution of~$K_i$, $i\in [\M]$, obtained from the state distribution $p_\rvecs$, the quantization mapping $q_\setQ(\cdot)$, and the target selection distributions~$\set{p_{O_k}}_{k=1}^\Ka$. For example, if $p_\rvecs$ and $p_{O_k}$ are the uniform distributions over $\setS$ and $[\Ma]$, respectively, and the quantization mapping preserves uniformity, we have that, for all $i \in [\M]$,
\begin{equation}
    p_{K_i}(k) = \sum_{m=0}^\Ma \Binpmf{m}{\Ma}{\frac{1}{\M}} \Binpmf{k}{\Ka}{\frac{m}{\Ma}}.
\label{eq:prior}
\end{equation}
Indeed, the number of targets having a given quantized state index $i \in [\M]$ follows a binomial distribution with parameters $(\Ma, 1/\M)$. Furthermore, given that this number is $m$, 
$K_i$ follows a binomial distribution with parameters $(\Ka, m/\M)$.

A soft decoder computes the marginal posterior probability 
\begin{equation}
    p_{K_i \given \rvecy}(k_i \given \vecy) = \!\!\sum_{\{k_j\}_{j\ne i} \in \set{0,1,\dots,\Ka}^{\M-1}} \!\!\!p_{\rveck \given \rvecy}([k_1,\dots,k_\M] \given \vecy) \label{eq:marginalization}
\end{equation}
for $i \in [\M]$, where
    $p_{\rveck \given \rvecy}(\veck \given \vecy) = \frac{p_{\rvecy \given \rveck}(\vecy \given \veck) p_\rveck(\veck)}{p_\rvecy(\vecy)} \propto p_{\rvecy \given \rveck}(\vecy \given \veck) p_\rveck(\veck)$ is the joint posterior probability of $\rveck$.
 The marginalization in~\eqref{eq:marginalization} is cumbersome. Therefore, we seek to decouple~\eqref{eq:received_signal_2} into $\M$ scalar Gaussian models of the form 
 \begin{equation}
     R_i = K_i + \normal(0,\xi_i), \quad i\in [\M]. \label{eq:scalar_model}
 \end{equation}
As a result, given $R_i = r_i$, the marginal posterior mean and variance of $K_i$ can be estimated by
\begin{align}
    \!\!\est{k}_i &= f(r_i,\xi_i) \define \frac{\sum_{k=0}^\Ka k p_{K_i}(k) \exp\mathopen{}\big(\!-\frac{(r_i-k)^2}{2\xi_i}\big)}{\sum_{k=0}^\Ka p_{K_i}(k) \exp\mathopen{}\big(\!-\frac{(r_i-k)^2}{2\xi_i}\big)}, \label{eq:posterior_mean}\\
    \!\!\est{v}_i &= g(r_i,\xi_i) \define \frac{\!\sum_{k=0}^\Ka (k-\est{k}_i)^2 p_{K_i}(k) \exp\mathopen{}\big(\!-\frac{(r_i-k)^2}{2\xi_i}\big)\!}{\sum_{k=0}^\Ka p_{K_i}(k) \exp\mathopen{}\big(\!-\frac{(r_i-k)^2}{2\xi_i}\big)},\! \label{eq:posterior_var}
\end{align}
respectively. 
The final estimate of $\rveck$ is obtained by rounding $\west{\veck} = \tp{[\est{k}_1,\dots,\est{k}_\M]}$ to the nearest integers. In the following, we present and compare three algorithms to obtain the scalar models that have been used in the literature~\cite{fengler2019sparcs,Meng2021,Qiao2024massive}. 

\subsection{Approximate Message Passing} \label{sec:AMP}
We first consider the \gls{AMP} estimator designed in~\cite[Sec.~IV]{fengler2019sparcs} for the \gls{UMA} setting. At each iteration $t$, the following updates are performed:
\begin{align}
    \west{\vecy}\of{t} &= \tp{\matC} \vecz\of{t-1} + \sqrt{\n\power} \west{\veck}\of{t-1}, \label{eq:AMP_effective_observation}\\
    \west{\veck}\of{t} &= f_{t}(\west{\vecy}\of{t}), \label{eq:AMP_denoise}\\
    \vecz\of{t} &= \vecy - \sqrt{\n\power} \matC \west{\veck}\of{t} + \frac{\M}{\n} \vecz\of{t-1} \langle f'_t(\west{\vecy}\of{t}) \rangle. \label{eq:residual}
\end{align}
Here, $\langle \cdot \rangle$ denotes the arithmetic mean, and 
$f_t(\vecr) \define \tp{[f(r_1,\xi\of{t-1}), \dots, f(r_\M,\xi\of{t-1})]}$ where $f(\cdot,\cdot)$ was defined in~\eqref{eq:posterior_mean} and where 
$\xi\of{t-1} \!=\! \vecnorm{\vecz\of{t-1}}^2/\n$. In~\eqref{eq:residual}, $f'_t$ is the componentwise derivative of $f_t$. The estimate $\west{\veck}$ in the denoising step~\eqref{eq:AMP_denoise} is given by the posterior mean of $\{K_i\}_{i=1}^\M$ obtained from the scalar models~\eqref{eq:scalar_model} with $\xi_i = \xi\of{t-1}, \forall i\in [\M]$, and the effective observation $\tp{[R_i,\dots,R_\M]} = \west{\vecy}\of{t}$. 

The complexity of each iteration is of order $O(\n \M)$. It can be reduced to $O(\max\{\M,\n\} \log \max\{\M,\n\})$ if~$\matC$ is 
a truncated Hadamard matrix, because $\matC \west{\veck}\of{t}$ and $\tp{\matC}\vecz\of{t-1}$ can be computed efficiently using the Hadamard transform~\cite{fengler2019sparcs,Amalladinne2022}. 

\subsection{Expectation Propagation} \label{sec:EP}

Next, we describe the \gls{EP} algorithm developed in
~\cite[Sec.~IV-D]{Meng2021}.
We write the posterior probability~$p_{\rveck \given \rvecy}$ as
\begin{align}
    p_{\rveck \given \rvecy}(\veck \given \vecy)
    &\propto 
     p_{\rvecy \given \rveck}(\vecy \given \veck) p_{\rveck}(\veck) \\ 
     &\approx 
      \normal(\vecy; \sqrt{\n\power}\matC \veck, \matidentity_\n) \textstyle\prod_{i=1}^\M p_{K_i}(k_i).
     \label{eq:approx_independent}
\end{align}
The approximation in~\eqref{eq:approx_independent} follows by ignoring the dependency between the $\set{K_i}_{i=1}^\M$. 
Following \gls{EP}, we write the posterior estimate 
in accordance with~\eqref{eq:approx_independent} as
    $\est{p}_{\rveck \given \rvecy}(\veck \given \vecy) = \est{p}_0(\veck) \prod_{i=1}^\M \est{p}_{i}(k_i)$, 
where
    $\est{p}_0(\veck) \propto \prod_{i=1}^\M \normal(k_i;\mu_{i0}, \xi_{i0})$  and 
    $\est{p}_{i}(k_i) \propto \normal(k_i;\mu_{i1}, \xi_{i1})$. 
We iteratively update $\set{\mu_{i0},\xi_{i0}}$, $\set{\mu_{i1}, \xi_{i1}}$,  and the approximate posterior mean $\est{k}_i$ and variance $\est{v}_i$.
Specifically, at iteration~$t$, $\est{k}_i\of{t} = f(\mu_{i0}\of{t},\xi_{i0}\of{t})$ and $\est{v}_i\of{t} = g(\mu_{i0}\of{t},\xi_{i0}\of{t})$. 
The terms $\set{\mu_{i1}, \xi_{i1}}$ are updated as 
    \begin{align}
        \xi_{i1}\of{t} &= (1/\est{v}_{i}\of{t} - 1/\xi_{i0}\of{t-1})^{-1}, \\\mu_{i1}\of{t} &= \xi_{i1}\of{t}(\est{k}_i\of{t}/ \est{v}_{i}\of{t} - \mu_{i0}\of{t-1}/\xi_{i0}\of{t-1}).
    \end{align}
    Furthermore, the terms $\set{\mu_{i0},\xi_{i0}}$ 
    are updated as
    \begin{align}
        \xi_{i0}\of{t} &= (1/\est{\xi}_{i0} - 1/\xi_{i1}\of{t})^{-1}, 
        \\
        \mu_{i0}\of{t} &= \xi_{i0}\of{t}(\est{\mu}_{i0} /\est{\xi}_{i0} - \mu_{i1}\of{t}/\xi_{i1}\of{t}).
    \end{align}
    Here, $\est{\xi}_{i0}$ is the $i$th diagonal element of the matrix
    \begin{equation} \label{eq:Xi0_EP}
        \west\matXi_0 = \matXi_1 - \matXi_1 \tp{\matC} ((\n\power)^{-1} \matidentity_\n + \matC\matXi_1\tp{\matC})^{-1} \matC \matXi_1 
    \end{equation}
    and $\est{\mu}_{i0}$ is the $i$th element of 
    \begin{equation}
        \est \vecmu_0 = \west\matXi_0(\matXi_1^{-1}\vecmu_1 + \sqrt{\n\power}\tp{\matC}\vecy)
    \end{equation}
    with $\vecmu_1 = \tp{\big[\mu_{11}\of{t}, \mu_{21}\of{t}, \dots, \mu_{\M1}\of{t}\big]}$ and $\matXi_{1} = \diag\big(\xi_{11}\of{t}, \xi_{21}\of{t},\dots,\xi_{\M1}\of{t}\big)$.  
    The complexity of each iteration is of order $O(\min\{\M,\n\}^2 \max\{\M,\n\})$, and is dominated by the computation of $\west{\matXi}_0$ in~\eqref{eq:Xi0_EP}. 

\subsection{Scalar \gls{AMP}} \label{sec:scalarAMP}
Finally, we consider the scalar \gls{AMP} algorithm derived in~\cite[Sec. IV-C]{Meng2021} and adopted in~\cite{Qiao2024massive} for a type-based estimation problem associated to federated learning. Note that this algorithm was derived as a simplification of \gls{EP} rather than of the \gls{AMP} algorithm in Section~\ref{sec:AMP}. In each iteration~$t$, 
the approximate posterior mean $\est{k}_i\of{t}$ and variance $\est{v}_i\of{t}$ of $K_i$ are given by $f(r_i\of{t},\xi_i\of{t})$ and $g(r_i\of{t},\xi_i\of{t})$, respectively. 
Here, the effective observation 
$r_i$ and effective noise variance 
$\xi_i$, $i\in [\M]$, are updated as
\begin{align}
    \xi_i\of{t} &= \bigg(\sum_{j=1}^\n \frac{c_{ji}^2}{(\n\power)^{-1} + v_j\of{t}}\bigg)^{-1}, \\
    r_i\of{t} &= \est{k}_i\of{t} + \xi_i\of{t} \sum_{j=1}^\n \frac{c_{ji}(y_j/\sqrt{\n\power} - z_j\of{t})}{(\n\power)^{-1} + v_j\of{t}},
\end{align}
where $c_{ji}$ is the $(j,i)$th entry of $\matC$. Furthermore, $v_j$ and $z_j$, $j\in [\n]$, are updated as
\begin{align}
    v_j\of{t} = \sum_{i=1}^\M c_{ji}^2 \est{v}_i\of{t},~
    z_j\of{t} = \sum_{i=1}^\M c_{ji} \est{k}_i\of{t} - v_j\of{t} \frac{y_j/\sqrt{\n\power} \!-\! z_j\of{t-1}}{(\n\power)^{-1} + v_j\of{t-1}}. \notag
\end{align}
The complexity of each iteration is of order $O(\n\M)$.

To summarize, \gls{AMP} has a lower complexity than \gls{EP}, and if a truncated-Hadamard codebook is used, it also has a lower complexity than scalar \gls{AMP}.
\section{Multi-Target Position Tracking} \label{sec:tracking}

We consider a scenario where $\Ka$ sensors track the position of $\Ma$ targets placed uniformly at random in a square area~$\setS$ of size $1 \times 1$. The area is divided into a regular grid consisting of $\M$ disjoint cells whose centroids form the set of quantized positions~$\setQ$. 
Each sensor tracks a target chosen uniformly at random from the $\Ma$ targets.\footnote{This is a simplistic model. More practical sensor deployment and target selection will be considered in future works.} 
Sensor~$k$ determines the position of its target, 
maps this position to the closest quantized position indexed by $W_k$, and sends the codeword~$\vecc_{W_k}$ to the receiver. 
In Fig.~\ref{fig:tracking}, we illustrate an example with $\Ma = 3$ targets, $\Ka = 4$ sensors, and $\M = 16$ quantized positions. 
\begin{figure}[t!]
    \centering
    \scalebox{.77}{
    \begin{tikzpicture}

        \draw[step=1.25cm, thin, dotted] (2.5+7.5,2.5) grid (7.5+7.5,7.5);
        \draw[] (2.5+7.5,7.5) -- (7.5+7.5,7.5) node[anchor=south east] {};
        \draw[] (7.5+7.5,2.5) -- (7.5+7.5,7.5) node[anchor=south east] {};
        \draw[] (2.5+7.5,2.5) -- (2.5+7.5,7.5) node[anchor=south east] {};
        \draw[] (2.5+7.5,2.5) -- (7.5+7.5,2.5) node[anchor=south east] {};

        \foreach \x in {1,2,3,4}
            \foreach \y in {1,2,3,4}
                \fill (0.625 + 1.25 + 7.5 + 1.25*\x,0.625 + 1.25 + 1.25*\y) circle[radius=1.5pt];


        \draw[dotted] (3.4+7.5,5.2) -- (3.1+7.5,4.9);
        \draw[dotted] (3.4+7.5,5.2) -- (2.5+7.5+1.25/2,2.5+1.25/2 + 1.25*2);
        \draw[dotted] (2.5+7.5+1.25/2,2.5+1.25/2 + 1.25*2) -- (3.1+7.5,4.9);
        \node[label={$\vecq_{W_1}$}] at (2.5+7.5+1.25/2,2.5+1.25/2 + 1.25*2 - 0.1) {};

        \draw[dotted] (5+7.5,6) -- (4.6+7.5,6.1);
        \draw[dotted] (5+7.5,6) -- (2.5+7.5+1.25/2+1.25,2.5+1.25/2 + 1.25*2);
        \draw[dotted] (2.5+7.5+1.25/2+1.25,2.5+1.25/2 + 1.25*2) -- (4.6+7.5,6.1);
        \node[label={$\vecq_{W_2}$}] at (2.5+7.5+1.25/2+1.25,2.5+1.25/2 + 1.25*2 - 0.6) {};

        \draw[dotted] (2.5+7.5+1.25/2+1.25*3,2.5+1.25/2) -- (5.3+1.25+7.5, 1+1.25+1.25);
        \draw[dotted] (4.9+1.25+7.5,0.8+2.5) -- (5.3+1.25+7.5, 1+1.25+1.25);
        \draw[dotted] (4.9+1.25+7.5,0.8+2.5) -- (2.5+7.5+1.25/2+1.25*3,2.5+1.25/2);
        
        \draw[dotted] (6.85+7.5,3.5) -- (5.3+1.25+7.5, 1+1.25+1.25);
        \draw[dotted] (6.85+7.5,3.5) -- (2.5+7.5+1.25/2+1.25*3,2.5+1.25/2);
        
        \node[label={$\vecq_{W_3} \!=\! \vecq_{W_4}$}] at (2.5+7.5+1.25/2+1.25*3-.2,2.5+1.25/2 - 0.7) {};

        \fill[red] (3.4+7.5,5.2) circle[radius=2pt];
        \node[label={[text=red, xshift=10pt, yshift=-12pt]$\rvecs_1$}] (M1) at (3.35+7.5,5.2){};

        \fill[red] (5+7.5,6) circle[radius=2pt];
        \node[label={[text=red, xshift=10pt, yshift=-12pt]$\rvecs_2$}] (M2)  at(4.95+7.5,6){};
        
        \fill[red] (5.3+1.25+7.5, 1+1.25+1.25) circle[radius=2pt];
        \node[label={[text=red, xshift=-2pt, yshift=0pt]$\rvecs_3$}] (MMa)  at(5.3+1.25+7.5, 1+1.25+1.25){};

        \draw[blue, very thick] (3+7.5,4.8) -- (3.2+7.5,5) node[anchor=south east] {};
        \draw[blue, very thick] (3+7.5,5) -- (3.2+7.5,4.8) node[label={[xshift=-12pt, yshift=-15pt] sensor $1$}] {};
        \begin{scope}[on background layer]
            \node[fill=gray!15, circle, minimum height=1cm] at (3.1+7.5,4.9){};
        \end{scope}

        \draw[blue, very thick] (4.5+7.5,6) -- (4.7+7.5,6.2) node[anchor=south east] {};
        \draw[blue, very thick] (4.5+7.5,6.2) -- (4.7+7.5,6) node[label={[xshift=-12pt, yshift=2pt]sensor $2$}] {};
        \begin{scope}[on background layer]
            \node[fill=gray!15, circle, minimum height=1cm] at (4.6+7.5,6.1){};
        \end{scope}

        \draw[blue, very thick] (4.8+1.25+7.5,1.25+0.7+1.25) -- (5+1.25+7.5,1.25+0.9+1.25) node[anchor=south east] {};
        \draw[blue, very thick] (4.8+1.25+7.5,1.25+0.9+1.25) -- (5+1.25+7.5,1.25+0.7+1.25) node[label={[xshift=-20pt, yshift=1pt] sensor $4$}] (SK) {};
        \begin{scope}[on background layer]
            \node[fill=gray!15, circle, minimum height=1cm] at (4.9+1.25+7.5,0.8+2.5){};
        \end{scope}

        \draw[blue, very thick] (6.75+7.5,3.4) -- (6.95+7.5,3.6) node[anchor=south east] {};
        \draw[blue, very thick] (6.75+7.5,3.6) -- (6.95+7.5,3.4) node[label={[xshift=18pt, yshift=-10pt]sensor $3$}] (SK_1) {};
        \begin{scope}[on background layer]
            \node[fill=gray!25, circle, minimum height=1cm] at (6.85+7.5,3.5){};
        \end{scope}




        \fill (15.3,5.5) circle[radius=1.5pt];
        \node[label={right:\text{Quantized positions}}] at(15.3,5.5){};

        \fill[red] (15.3,5) circle[radius=2pt];
        \node[label={right:\text{Targets}}] at(15.3,5){};

        \draw[blue, very thick] (15.2,4.4) -- (15.4,4.6) node[anchor=east] {};
        \draw[blue, very thick] (15.2,4.6) -- (15.4,4.4) node[anchor=south east] {};
        \node[label={right:\text{Sensors}}] at(15.3,4.5){};
    \end{tikzpicture}
    }
    \vspace{-.1cm}
    \caption{An example of multi-target position tracking in a square with $\Ma = 3$ targets, $\Ka = 4$ sensors, and $\M = 16$ quantized positions. 
    }
    \label{fig:tracking}
    \vspace{-.3cm}
\end{figure}


We generate 
$\matC$ as a truncated Hadamard matrix. We assume that the receiver uses the \gls{AMP}, scalar \gls{AMP}, or \gls{EP} decoder described in Section~\ref{sec:decoder}, with the prior given in~\eqref{eq:prior}.\footnote{This requires that the receiver knows both $\Ka$ and $\Ma$. Our decoder can be extended to the case of unknown $\Ka$ and $\Ma$. In this case, we initialize these parameters and refine their values, and also the prior, along the iterations.} We set the maximum number of iterations to $10$, and terminate the iterations early if the estimate of~$\rveck$ remains unchanged after an iteration. We first compare the ability of these decoders to recover $\rveck$ by evaluating the average total variation distance $\dTV$ in~\eqref{eq:TV}.
In Fig.~\ref{fig:TV_vs_Ma}, we plot $\dTV$ for the three decoders as a function of the number of targets $\Ma$ for a setting with $\n \in \set{250, 500}$ channel uses, $\Ka = 50$ sensors, $\log_2 \M = 10$ bits, and the \gls{SNR} $\power = -12\dB$. For small $\Ma$ values, the three decoders perform similarly. However, as $\Ma$ increases, \gls{AMP} outperforms both \gls{EP} and scalar \gls{AMP}, which exhibits the worst performance. 
Overall, \gls{AMP} is preferable because of both complexity and performance advantages. 



\begin{figure}[t!]
    \centering
    \vspace{-.1cm}
\begin{tikzpicture}
    \tikzstyle{every node}=[font=\footnotesize] 
    \begin{semilogyaxis}[
    scale only axis,
    width=2.7in,
    height=1.1in,
    grid=both,
    grid style={line width=.1pt, draw=gray!10},
    major grid style={line width=.2pt,draw=gray!50},
    xmin=5,
    xmax=50,
    extra x ticks={5},
    xlabel={number of targets, $\Ma$},
    ymin=1e-4,
    ymax=1,
    ytick = {1e-4,1e-3,1e-2,1e-1,1e0},
    ylabel={average total variation, $\dTV$},
     ylabel style = {xshift=-1mm},
     legend style={draw=none,opacity=.9,at={(0.98,0.01)},anchor=south east},
    ]
        
       \addplot [color=red,line width=1.2pt]
        table[row sep=crcr]{%
            5.0e+00 3.0597e-04 \\ 
1.0e+01 1.5910e-03 \\ 
1.5e+01 3.7705e-03 \\ 
2.0e+01 5.7143e-03 \\ 
2.5e+01 7.2439e-03 \\ 
3.0e+01 8.6063e-03 \\ 
3.5e+01 1.0260e-02 \\ 
4.0e+01 1.0624e-02 \\ 
4.5e+01 1.1284e-02 \\ 
5.0e+01 1.2129e-02 \\ 
        };
        \addlegendentry{AMP};
        
  \addplot [color=black,dashed,line width=1.2pt]
        table[row sep=crcr]{%
           5.0e+00 4.1545e-04 \\ 
1.0e+01 1.5833e-03 \\ 
1.5e+01 3.5436e-03 \\ 
2.0e+01 5.3978e-03 \\ 
2.5e+01 9.8679e-03 \\ 
3.0e+01 2.2539e-02 \\ 
3.5e+01 8.1945e-02 \\ 
4.0e+01 1.6663e-01 \\ 
4.5e+01 2.7250e-01 \\ 
5.0e+01 3.2896e-01 \\ 
            };
        \addlegendentry{scalar AMP};

                \addplot [color=blue,mark=*,only marks]
        table[row sep=crcr]{%
            5.0e+00 3.9550e-04 \\ 
1.0e+01 1.5498e-03 \\ 
1.5e+01 3.4869e-03 \\ 
2.0e+01 5.2949e-03 \\ 
2.5e+01 7.2925e-03 \\ 
3.0e+01 8.5551e-03 \\ 
3.5e+01 9.2116e-03 \\ 
4.0e+01 1.0570e-02 \\ 
4.5e+01 1.1622e-02 \\ 
5.0e+01 1.1941e-02 \\ 
                };
        \addlegendentry{EP};

        \addplot [color=blue,mark=*,only marks]
        table[row sep=crcr]{%
            5.0e+00 3.5971e-03 \\ 
1.0e+01 1.0763e-02 \\ 
1.5e+01 2.3562e-02 \\ 
2.0e+01 4.0813e-02 \\ 
2.5e+01 6.0493e-02 \\ 
3.0e+01 8.0704e-02 \\ 
3.5e+01 1.1055e-01 \\ 
4.0e+01 1.4699e-01 \\ 
4.5e+01 1.7488e-01 \\ 
5.0e+01 2.2887e-01 \\ 
                };
        
       \addplot [color=red,line width=1.2pt,forget plot]
        table[row sep=crcr]{%
            5.0e+00 3.6314e-03 \\ 
1.0e+01 1.0406e-02 \\ 
1.5e+01 2.4462e-02 \\ 
2.0e+01 4.1467e-02 \\ 
2.5e+01 5.7090e-02 \\ 
3.0e+01 7.3804e-02 \\ 
3.5e+01 8.8535e-02 \\ 
4.0e+01 1.0660e-01 \\ 
4.5e+01 1.2245e-01 \\ 
5.0e+01 1.3226e-01 \\ 
        };
        
  \addplot [color=black,dashed,line width=1.2pt, forget plot]
        table[row sep=crcr]{%
           5.0e+00 3.4469e-03 \\ 
1.0e+01 1.0731e-02 \\ 
1.5e+01 2.7437e-02 \\ 
2.0e+01 1.5388e-01 \\ 
2.5e+01 3.8328e-01 \\ 
3.0e+01 6.4350e-01 \\ 
3.5e+01 7.7769e-01 \\ 
4.0e+01 8.6432e-01 \\ 
4.5e+01 9.1856e-01 \\ 
5.0e+01 9.3964e-01 \\ 
            };

        \coordinate (Center) at (axis cs:15,2.1e-2);
   		\coordinate (Radius) at (axis cs:15.7,4e-2);
   		
   		\coordinate (Center2) at (axis cs:15,3.5e-3);
   		\coordinate (Radius2) at (axis cs:15.7,7e-3);

        \node[align = center] at (axis cs:16,1.4e-3) () {$\n = 500$};
        \node[align = center] at (axis cs:12,6.3e-2) () {$\n = 250$};
        
      \end{semilogyaxis}
      \pgfextractx{\rx}{\pgfpointdiff{\pgfpointanchor{Radius}{center}}{\pgfpointanchor{Center}{center}}}%
	\pgfextracty{\ry}{\pgfpointdiff{\pgfpointanchor{Radius}{center}}{\pgfpointanchor{Center}{center}}}%
	\draw (Center) ellipse [x radius = \rx, y radius = \ry]; 
	
	\pgfextractx{\rx}{\pgfpointdiff{\pgfpointanchor{Radius2}{center}}{\pgfpointanchor{Center2}{center}}}%
	\pgfextracty{\ry}{\pgfpointdiff{\pgfpointanchor{Radius2}{center}}{\pgfpointanchor{Center2}{center}}}%
	\draw (Center2) ellipse [x radius = \rx, y radius = \ry]; 
    \end{tikzpicture}
    \vspace{-1.2cm}
    \caption{The average total variation $\dTV$ vs. the number of targets $\Ma$ for 
    $\Ka = 50$ sensors, $\log_2\M = 10$ bits, and the \gls{SNR} $\power = -12\dB$.}
    \label{fig:TV_vs_Ma}
    \vspace{-.5cm}
\end{figure}


Next, we investigate the impact of the quantization resolution, represented by the number of bits $\log_2\M$. In Fig.~\ref{fig:TV_WS_vs_B}, we plot both $\dTV$ and the average Wasserstein distance $\dWS_2$ achieved with \gls{AMP} as functions of $\log_2 \M$ for $\n \in \set{500,1000,2000}$ channel uses, $\Ka = 100$ sensors, $\Ma = 10$ targets, and the \gls{SNR} $P = -27$ dB. We also show the value of $\dWS_2$ achieved if there is no communication error (i.e., if $\west{\rveck} = \rveck$), which accounts only for the quantization distortion. As the quantization resolution increases, this distortion decreases. However, increasing~$\M$ for a fixed $\n$ leads to a higher transmission rate, and a consequent degradation of the communication performance, as shown by the increase of $\dTV$ observed in Fig.~\ref{fig:TV_WS_vs_B}(a). Due to this tradeoff, the average Wasserstein distance is minimized at an intermediate value of $\M$, as observed in Fig.~\ref{fig:TV_WS_vs_B}(b). The optimal value of $\M$ exceeds $\n$, indicating that the use of an orthogonal codebook, as in~\gls{TBMA}, is suboptimal. As $\n$ grows, the $\dWS_2$ curve flattens out around its minimum. 
\begin{figure}[t!]
    \centering
    \hspace{-5mm}
\subfigure[total variation]{ \label{fig:TV_vs_B}
\begin{tikzpicture}
    \tikzstyle{every node}=[font=\footnotesize] 
    \begin{semilogyaxis}[
    scale only axis,
    width=1.2in,
    height=1.4in,
    grid=both,
    grid style={line width=.1pt, draw=gray!10},
    major grid style={line width=.2pt,draw=gray!50},
    xmin=4,
    xmax=18,
    xtick={4,8,12,16,18},
    xlabel={number of bits, $\log_2 \M$},
    ymin=1e-2,
    ymax=1e-1,
    yminorticks=true,
    ylabel style = {yshift=-3mm},
    yticklabel style = {xshift=1mm},
    ylabel={average total variation, \dTV},
     legend pos=north west,
    ]

        \addplot [color=red,line width=1.2pt]
        table[row sep=crcr]{%
            2.0e+00 1.0051e-02 \\ 
4.0e+00 2.1202e-02 \\ 
6.0e+00 2.5906e-02 \\ 
8.0e+00 2.7894e-02 \\ 
1.0e+01 2.8796e-02 \\ 
1.2e+01 2.9703e-02 \\ 
1.4e+01 2.9852e-02 \\ 
1.6e+01 3.0126e-02 \\ 
18 0.0308\\
            };

        
       \addplot [color=blue,line width=1.2pt]
        table[row sep=crcr]{%
           2.0e+00 6.7365e-03 \\ 
4.0e+00 1.4252e-02 \\ 
6.0e+00 1.7543e-02 \\ 
8.0e+00 1.8512e-02 \\ 
1.0e+01 1.8856e-02 \\ 
1.2e+01 1.9131e-02 \\ 
1.4e+01 1.9161e-02 \\ 
1.6e+01 1.9092e-02 \\ 
18 0.019229976634953 \\
            };

        \addplot [color=black,line width=1.2pt]
        table[row sep=crcr]{%
           2.0e+00 1.3574e-02 \\ 
4.0e+00 3.0124e-02 \\ 
6.0e+00 3.7743e-02 \\ 
8.0e+00 4.1015e-02 \\ 
1.0e+01 4.3472e-02 \\ 
1.2e+01 4.6718e-02 \\ 
1.4e+01 4.9186e-02 \\ 
1.6e+01 5.3422e-02 \\ 
18 0.0573\\
            };

        \node[align = center,color=red] at (axis cs:14.7,3.5e-2) () {$\n = 1000$};
        \node[align = center,color=black] at (axis cs:14.7,6.5e-2) () {$\n = 500$};
        \node[align = center,color=blue] at (axis cs:14.7,1.7e-2) () {$\n = 2000$};
      \end{semilogyaxis}
    \end{tikzpicture}
    }
    \hspace{-6mm}
\subfigure[Wasserstein distance]{
    \begin{tikzpicture}
    \tikzstyle{every node}=[font=\footnotesize] 
    \begin{axis}[
    scale only axis,
    width=1.3in,
    height=1.4in,
    grid=both,
    grid style={line width=.1pt, draw=gray!10},
    major grid style={line width=.2pt,draw=gray!50},
    xmin=4,
    xmax=18,
    xtick={4,8,12,16,18},
    xlabel={number of bits, $\log_2 \M$},
    ymin=0,
    ymax=.1,
    ytick={0,.02,.04,.06,.08,.1},
    yticklabels={$0$,$0.02$,$0.04$,$0.06$,$0.08$,$0.1$},
    ylabel={average Wasserstein distance, $\dWS_2$},
    ylabel style={yshift=-2mm, xshift=-1mm}, %
    yticklabel style = {xshift=.5mm},
     legend style={at={(.15,0.99)}, anchor=north west, legend cell align=left, align=left, fill=white,draw=black}
    ]
        \addplot [color=gray!70,line width=2.5pt]
        table[row sep=crcr]{%
             4.0e+00 1.0146e-01 \\ 
6.0e+00 5.0627e-02 \\ 
8.0e+00 2.5401e-02 \\ 
1.0e+01 1.2702e-02 \\ 
1.2e+01 6.3411e-03 \\ 
1.4e+01 3.1730e-03 \\ 
1.6e+01 1.5864e-03 \\ 
1.8e+01 7.937e-04 \\ 
            }
            node [pos=.7,pin={[align=center,text=black,pin edge={solid,black,<-},pin distance=2.5mm]90:perfect \\ communication},inner sep=1pt] {};
            ;
        
        \addplot [color=red,line width=1.2pt]
        table[row sep=crcr]{%
            2.0e+00 2.0630e-01 \\ 
4.0e+00 1.1599e-01 \\ 
6.0e+00 7.8868e-02 \\ 
8.0e+00 6.6943e-02 \\ 
1.0e+01 6.4470e-02 \\ 
1.2e+01 6.3623e-02 \\ 
1.4e+01 6.3409e-02 \\ 
1.6e+01 6.3540e-02 \\ 
1.8e+01 0.0647 \\
            };

        
       \addplot [color=blue,line width=1.2pt]
        table[row sep=crcr]{%
            2.0e+00 2.0570e-01 \\ 
4.0e+00 1.1202e-01 \\ 
6.0e+00 7.1957e-02 \\ 
8.0e+00 5.7433e-02 \\ 
1.0e+01 5.2830e-02 \\ 
1.2e+01 5.2083e-02 \\ 
1.4e+01 5.1541e-02 \\ 
1.6e+01 5.1562e-02 \\ 
18 0.051602545957150 \\
            };

        \addplot [color=black,line width=1.2pt]
        table[row sep=crcr]{%
            2.0e+00 2.0742e-01 \\ 
4.0e+00 1.2068e-01 \\ 
6.0e+00 8.7468e-02 \\ 
8.0e+00 7.7856e-02 \\ 
1.0e+01 7.6858e-02 \\ 
1.2e+01 7.8321e-02 \\ 
1.4e+01 8.1461e-02 \\ 
1.6e+01 8.3622e-02 \\ 
1.8e+01 0.0885 \\ 
            };

        \node[align = center,color=blue] at (axis cs:14.8,0.047) () {$\n = 2000$};
        \node[align = center,color=red] at (axis cs:14.8,0.07) () {$\n = 1000$};
        \node[align = center,color=black] at (axis cs:14.8,0.09) () {$\n = 500$};
      \end{axis}
    \end{tikzpicture}
    }
    \hspace{-5mm}
    \vspace{-.6cm}
    \caption{The average total variation $\dTV$ and average $2$-Wasserstein distance $\dWS_2$ vs. the number of bits $\log_2\M$ for 
    $\Ka = 100$ sensors, $\Ma = 10$ targets, and the \gls{SNR} $P = -27\dB$.}
    \label{fig:TV_WS_vs_B}
    \vspace{-.4cm}
\end{figure}

For the considered coding scheme, $\n$ and $\M$ cannot be increased beyond the values chosen in Fig.~\ref{fig:TV_WS_vs_B} because of complexity. To address the case $\log_2 \M \gg 18$, 
one needs to use coded-compressed sensing solutions as in~\cite{fengler2019sparcs,Amalladinne2022}. 

\section{Conclusions} \label{sec:conclusion}
We formulated \gls{TUMA}: a framework in which a large number of users track the state of multiple processes. Each user quantizes the state of its tracked process and sends, in an uncoordinated manner, the quantization index to the receiver, which then estimates the type of the quantized states. We used the Wasserstein distance to evaluate the type estimation error. Our results showcased the effectiveness of \gls{AMP} for accurate type estimation. Furthermore, we identified a trade-off between quantization resolution and communication error when minimizing the Wasserstein distance.


\appendices

\bibliographystyle{IEEEtran}
\bibliography{IEEEabrv,./references}

\end{document}